\documentclass[aps,pre,twocolumn,showpacs,superscriptaddress]{revtex4-1}
\usepackage{epsfig}
\usepackage{times}

\begin{document}

\title{Correlation of positive and negative reciprocity fails to confer an evolutionary advantage:\\Phase transitions to elementary strategies}

\author{Attila Szolnoki}
\affiliation{Institute of Technical Physics and Materials Science, Research Centre for Natural Sciences, Hungarian Academy of Sciences, P.O. Box 49, H-1525 Budapest, Hungary}

\author{Matja{\v z} Perc}
\affiliation{Faculty of Natural Sciences and Mathematics, University of Maribor, Koro{\v s}ka cesta 160, SI-2000 Maribor, Slovenia}

\begin{abstract}
Economic experiments reveal that humans value cooperation and fairness. Punishing unfair behavior is therefore common, and according to the theory of strong reciprocity, it is also directly related to rewarding cooperative behavior. However, empirical data fail to confirm that positive and negative reciprocity are correlated. Inspired by this disagreement, we determine whether the combined application of reward and punishment is evolutionary advantageous. We study a spatial public goods game, where in addition to the three elementary strategies of defection, rewarding and punishment, a fourth strategy combining the later two competes for space. We find rich dynamical behavior that gives rise to intricate phase diagrams where continuous and discontinuous phase transitions occur in succession. Indirect territorial competition, spontaneous emergence of cyclic dominance, as well as divergent fluctuations of oscillations that terminate in an absorbing phase are observed. Yet despite the high complexity of solutions, the combined strategy can survive only in very narrow and unrealistic parameter regions. Elementary strategies, either in pure or mixed phases, are much more common and likely to prevail. Our results highlight the importance of patterns and structure in human cooperation, which should be considered in future experiments.
\end{abstract}

\pacs{87.23.Ge, 89.75.Fb}
\maketitle

\section{Introduction}

Humans have mastered the art of cooperation like no other species \cite{nowak_11, apicella_n12}. Regardless of kinship and individual loss, we work together to achieve feats that are impossible to achieve alone. We have developed a very keen sense of fairness to uphold cooperative behavior in our societies \cite{fehr_qje99, henrich_aer01}, and we frequently punish those that do not cooperate in the pursuit of personal benefits and elevated status \cite{fehr_aer00, henrich_s06b, sigmund_tee07}. There also exist evidence for common marmosets and chimpanzees to show similar preferences regarding altruism and reward division \cite{burkart_pnas07, proctor_pnas13}, suggesting a long evolutionary history to the human sense of fairness. Although the origins of this behavior are not fully understood, there exist evidence for between-group conflicts \cite{bowles_11} and the provisioning for someone else's young \cite{hrdy_11} as viable for igniting the evolution of remarkable other-regarding abilities of the genus \textit{Homo}.

Like the origins of cooperative behavior, so too its later development and evolution continue to intrigue and stimulate new research across social and natural sciences \cite{szabo_pr07, perc_bs10, perc_jrsi13, nowak_jtb12, rand_tcs13}. Although key mechanisms have been identified that promote the evolution of cooperation \cite{nowak_s06}, there is still disagreement between theory and experiment on many key issues. Two examples have recently attracted notable interest. The first concerns network reciprocity \cite{nowak_n92b, hauert_n04, santos_prl05, fu_prsb11}, according to which cooperators are able to exploit the structure of interaction networks to offset inherent evolutionary disadvantages over defectors. Recent large-scale human experiments, however, fail to provide evidence in support of network reciprocity \cite{gracia-lazaro_pnas12}. The second example is of direct relevance for the present work, and it concerns the strong reciprocity model \cite{gintis_jtb00, fehr_hn02, boyd_pnas03, bowles_tpb04}. The later postulates that positive and negative reciprocity are directly correlated. In theory, it indeed seems reasonable to assume that rewarding cooperative behavior and punishing unfair behavior are to be seen as the two sides of the same preference for fairness. Yet recently gathered empirical data suggest otherwise \cite{yamagishi_pnas12, egloff_pnas13}. In fact, Yamagishi et al. \cite{yamagishi_pnas12} have performed a series of experiments and concluded that there is no correlation between the tendencies to reject unfair offers in the ultimatum game \cite{guth_jebo82} and the tendencies to exhibit prosocial behavior in other games \cite{rand_s09, rand_pnas11}. Moreover, the analysis of private household data from the Socio-Economic Panel of the German Institute for Economics Research presented by Egloff et al. \cite{egloff_pnas13} has revealed that positive and negative reciprocity vary independently of each other, thus providing a severe challenge to the strong reciprocity model of the evolution of human cooperation. While the rejection of unfair offers, which ought to be seen equivalent to punishing defection \cite{sigmund_sa02}, is simply a tacit strategy for avoiding the imposition of an inferior status, the act of cooperating appears to have an altogether different motivational background.

The described disagreement between the strong reciprocity model and empirical data invites an interdisciplinary approach, which promises to shed light on the subject from a different perspective. In the present paper, we therefore apply evolutionary game theory \cite{sigmund_93, weibull_95, hofbauer_98, nowak_06, sigmund_10} and methods of statistical physics \cite{binder_88, liggett_85} to determine whether there are evolutionary advantages to be gained by adopting a strategy that punishes defectors as well as rewards cooperators, as opposed to doing just one or the other. While the elementary strategies of rewarding and punishment have received ample attention in the recent past \cite{hilbe_prsb10, sigmund_n10, rand_nc11, hilbe_srep12, perc_njp12, szolnoki_njp12}, little is known about their combined effectiveness. To amend this, we propose and study a modified spatial public goods game \cite{brandt_prsb03, szolnoki_pre09c}, where defectors compete with cooperators that punish defectors, reward other cooperators, as well as do both. We intentionally leave out cooperators that neither reward nor punish in order to avoid the second-order free-riding problem \cite{fehr_n04, fowler_n05b}, and to thus be able to focus solely on the effectiveness of the combined strategy against the three elementary strategies of defection, rewarding and punishment.

As we will show in what follows, although the spatiotemporal dynamics of the evolutionary game is very complex and interesting from the physics point of view, there exist only narrow and realistically unlikely parameter regions where the combined strategy is able to survive. Given the lack of notable evolutionary advantages of correlating positive and negative reciprocity, the outcome of the experiments by Yamagishi et al. \cite{yamagishi_pnas12} and Egloff et al. \cite{egloff_pnas13} can thus be better understood, though the complexity of solutions also lends some support to the strong reciprocity hypothesis as being viable at least under certain special circumstances. We will present compelling evidence to support these conclusions in section III, while in the next section we first describe the studied spatial public goods game and the methods in more detail.

\section{Public goods game with positive and negative reciprocity}

As a frequently used paradigm of social conflicts and human cooperation, the public goods game is staged on a square lattice with periodic boundary conditions where $L^2$ players are arranged into overlapping groups of size $G=5$ such that everyone is connected to its $G-1$ nearest neighbors. Accordingly, each individual belongs to $g=1,\ldots,G$ different groups. The square lattice is the simplest of networks that allows us to go beyond the unrealistic well-mixed population assumption, and as such it allows us to take into account the fact that the interactions among humans are inherently structured rather than random. By using the square lattice, we also continue a long-standing history that begun with the work of Nowak and May \cite{nowak_n92b}, who were the first to show that the most striking differences in the outcome of an evolutionary game emerge when the assumption of a well-mixed population is abandoned for the usage of a structured population. Many have since followed the same practice \cite{brandt_prsb03, nakamaru_eer05, helbing_njp10} (for a review see \cite{perc_jrsi13}), and there exist ample evidence in support of the claim that, especially for games that are governed by group interactions \cite{szolnoki_pre09c, szolnoki_pre11c}, using the square lattice suffices to reveal all the feasible evolutionary outcomes, and also that these are qualitatively independent of the interaction structure.

Initially each player on site $x$ is designated either as a defector ($s_x = D$), cooperator that punishes defectors ($s_x = P$), cooperator that rewards other cooperators ($s_x = R$), or cooperator that both punishes defectors as well as rewards other cooperators ($s_x = B$) with equal probability. All three cooperative strategies ($P$, $R$ and $B$) contribute a fixed amount (here considered being equal to $1$ without loss of generality) to the public good while defectors contribute nothing. The sum of all contributions in each group is multiplied by the synergy factor $r$ and the resulting public goods are distributed equally amongst all the group members irrespective of their strategies. In addition, a defector suffers a fine $\beta/(G-1)$ from each punisher ($P$ or $B$) within the interaction neighborhood, which in turn requires the punisher to bear the cost $\gamma/(G-1)$ on each defecting individual in the group. A defector thus suffers the maximal fine $\beta$ if it is surrounded solely by punishers, while a lonely punisher bears the largest cost $\gamma$ if it is surrounded solely by defectors. Similarly, every cooperator is given the reward $\beta/(G-1)$ from every $R$ and $B$ player within the group, while each of them has to bear the cost of rewarding $\gamma/(G-1)$ for every cooperator that is rewarded. As a technical comment, we note that the application of payoffs normalized by $G-1$ enables relevant comparisons with the evolutionary outcomes on other interaction networks where players might differ in their degree and group size. Moreover, we use an equally strong fine and reward at the same cost, technically the same pair of $(\beta, \gamma)$ values for reward and punishment, which ensures a fair evaluation of the evolutionary advantage of both strategies. By decoupling these parameters, for example by administering high fines and low rewards at the same cost to both punishers and those that reward, would confer an unfair advantage to punishment because it would then be relatively less costly than rewarding. Since giving equal chances for success is of paramount importance for assessing evolutionary viability, we do not decouple $\beta$ and $\gamma$ for reward and punishment, and we also award limitless resources to all competing strategies.

In agreement with the described rules of the game, the payoff values of the four competing strategies obtained from each group $g$ are thus:
\begin{widetext}
\begin{eqnarray}
\Pi_D^{g} &=& r(N_P+N_R+N_B)/G - \beta (N_P+N_B) /(G-1),\nonumber \\
\Pi_P^{g} &=& r(N_P+N_R+N_B+1)/G - \gamma (N_D) /(G-1) + \beta (N_R+N_B) / (G-1), \nonumber\\
\Pi_R^{g} &=& r(N_P+N_R+N_B+1)/G - \gamma (N_P+N_R+N_B) /(G-1)+ \beta (N_R+N_B)/(G-1), \nonumber\\
\Pi_B^{g} &=& r(N_P+N_R+N_B+1)/G - \gamma + \beta (N_R+N_B) / (G-1), \nonumber
\label{payoff}
\end{eqnarray}
\end{widetext}
where $N_{s_x}$ denotes the number of other players with strategy $s_x$ in the group.

Monte Carlo simulations of the public goods game are carried out comprising the following elementary steps. A randomly selected player $x$ plays the public goods game with its $G-1$ partners as a member of all the $g=1,\ldots,G$ groups, whereby its overall payoff $\Pi_{s_x}$ is thus the sum of all the payoffs $\Pi_{s_x}^{g}$ acquired in each individual group. Next, player $x$ chooses one of its nearest neighbors at random, and the chosen co-player $y$ also acquires its payoff $\Pi_{s_y}$ in the same way. Finally, player $x$ enforces its strategy $s_x$ onto player $y$ with a probability given by the Fermi function $w(s_x \to s_y)=1/\{1+\exp[(\Pi_{s_y}-\Pi_{s_x}) /K]\}$, where $K=0.5$ quantifies the uncertainty by strategy adoptions \cite{szolnoki_pre09c}, implying that better performing players are readily adopted, although it is not impossible to adopt the strategy of a player performing worse. Such errors in decision making can be attributed to mistakes and external influences that adversely affect the evaluation of the opponent. Each Monte Carlo step (MCS) gives a chance for every player to enforce its strategy onto one of the neighbors once on average.

The average fractions of defectors ($\rho_{D}$), cooperators that punish ($\rho_{P}$), cooperators that reward ($\rho_{R}$), and cooperators that do both ($\rho_{B}$) on the square lattice were determined in the stationary state after a sufficiently long relaxation time. Depending on the actual conditions (proximity to phase transition points and the typical size of emerging spatial patterns), the linear system size was varied from $L=400$ to $7200$ and the relaxation time was varied from $10^4$ to $10^5$ MCS to ensure that the statistical error is comparable with the line thickness in the figures. We note that the random initial state may not necessarily yield a relaxation to the most stable solution of the game even at such a large system size ($L=7200$). To verify the stability of different solutions, we have therefore applied prepared initial states (see Fig.~10 in \cite{szolnoki_pre11b}), and we have followed the same procedure as described previously in \cite{szolnoki_pre11}. Next we proceed with presenting the main results.

\begin{figure}
\centerline{\epsfig{file=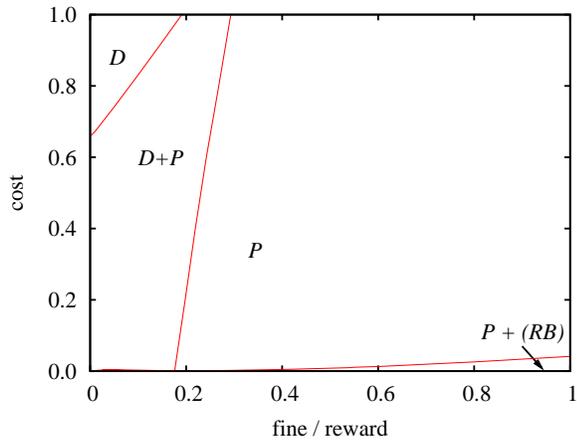,width=8.5cm}}
\caption{\label{phd_r4_5} Full $\beta-\gamma$ phase diagram, as obtained for $r=4.5$. Solid red lines denote continuous phase transitions. If defectors die out strategies $R$ and $B$ become equivalent (see main text for details), hence the $(RB)$ notation in the lower right corner of the parameter plane. The vertical resolution hides the intricate structure of the phase diagram for very low values of $\gamma$, which we therefore show separately in Fig.~\ref{phd_r4_5_enlarged}.}
\end{figure}

\section{Results}

\begin{figure}
\centerline{\epsfig{file=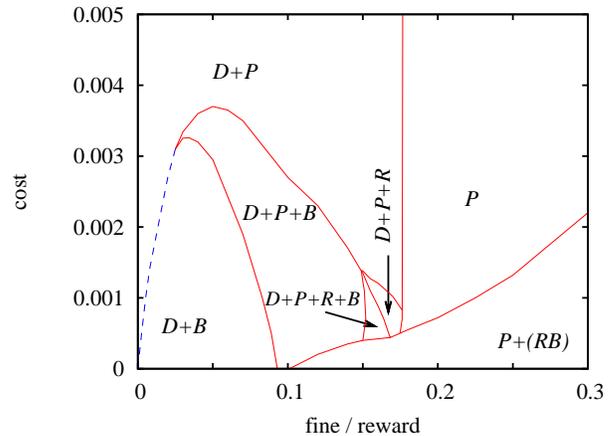,width=8.8cm}}
\caption{\label{phd_r4_5_enlarged} Enlarged part of the full $\beta-\gamma$ phase diagram depicted in Fig.~\ref{phd_r4_5}, zooming in on the very small ($\sim 10^3$ smaller than $\beta$) values of $\gamma$. Solid red lines denote continuous phase transitions, while the blue dashed line denotes discontinuous phase transitions. The later are due to indirect territorial competition between strategies $P$ and $B$ who compete independently against the defectors. A representative cross-section of the phase diagrams is presented in Fig.~\ref{r4_5_cross}.}
\end{figure}

\begin{figure}[b]
\centerline{\epsfig{file=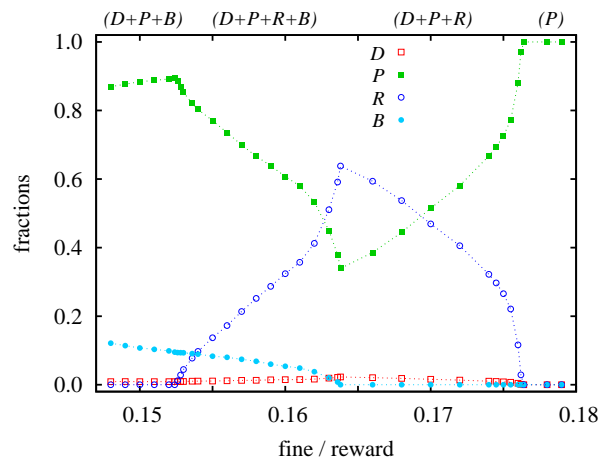,width=8.5cm}}
\caption{\label{r4_5_cross} Cross-section of the phase diagram depicted in Fig.~\ref{phd_r4_5_enlarged}, as obtained for $\gamma=0.0007$. Depicted are stationary fractions of the four competing strategies in dependence on $\beta$. Stable solutions are denoted along the top axis. Unlike the $D+P \to D+B$ phase transition denoted dashed blue in Fig.~\ref{phd_r4_5_enlarged}, in this cross-section all phase transitions are continuous.}
\end{figure}

\begin{figure*}
\centerline{\epsfig{file=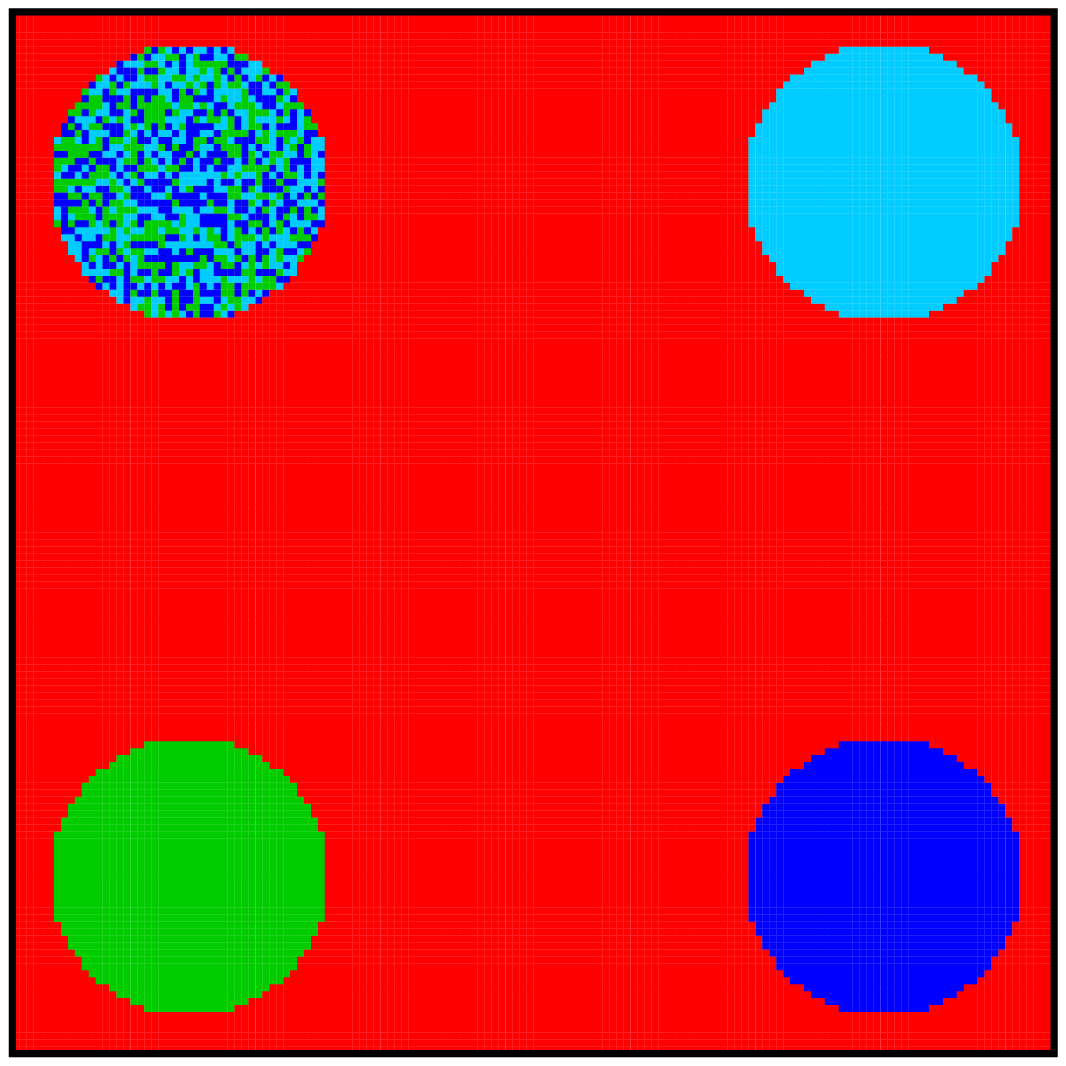,width=4.6cm}\epsfig{file=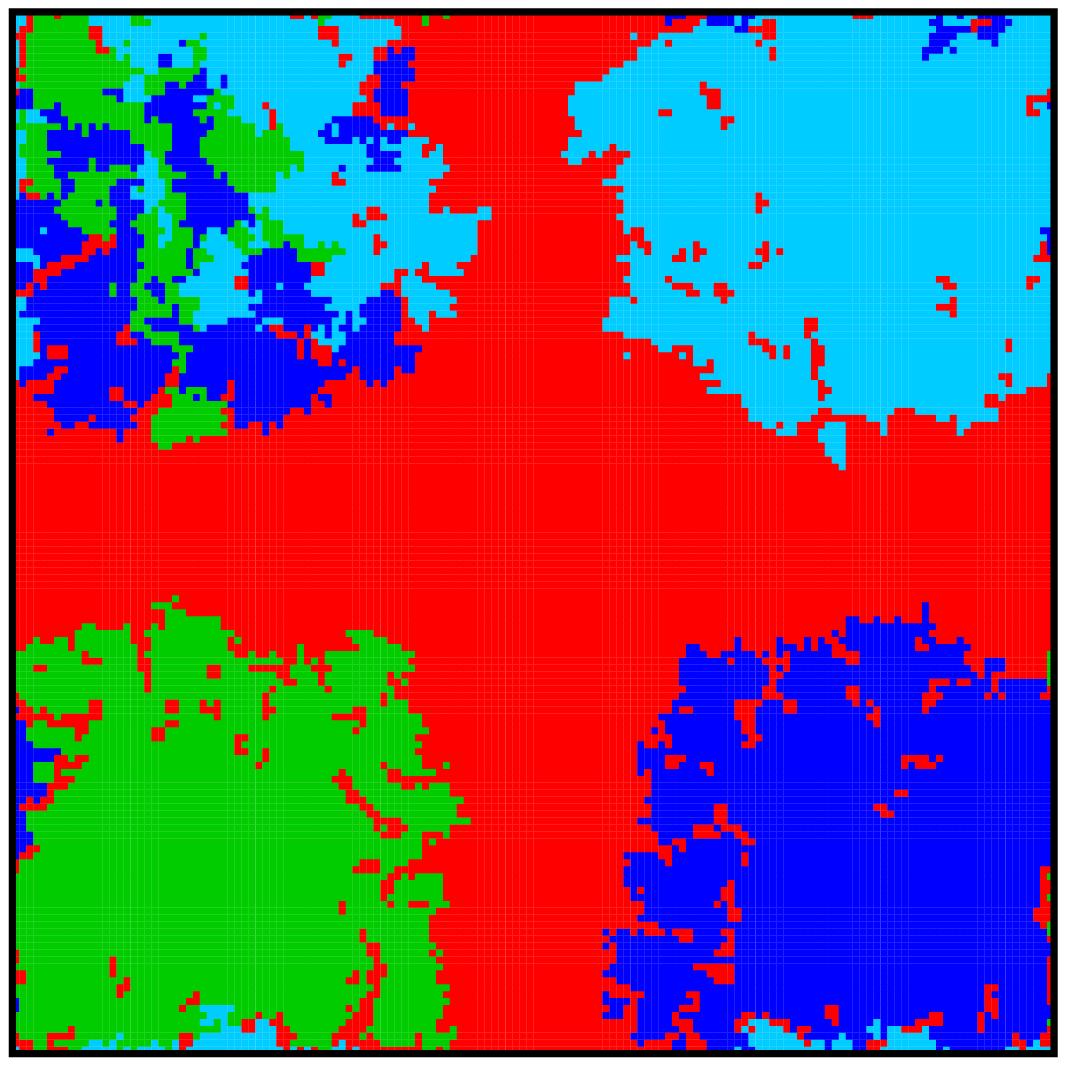,width=4.6cm}\epsfig{file=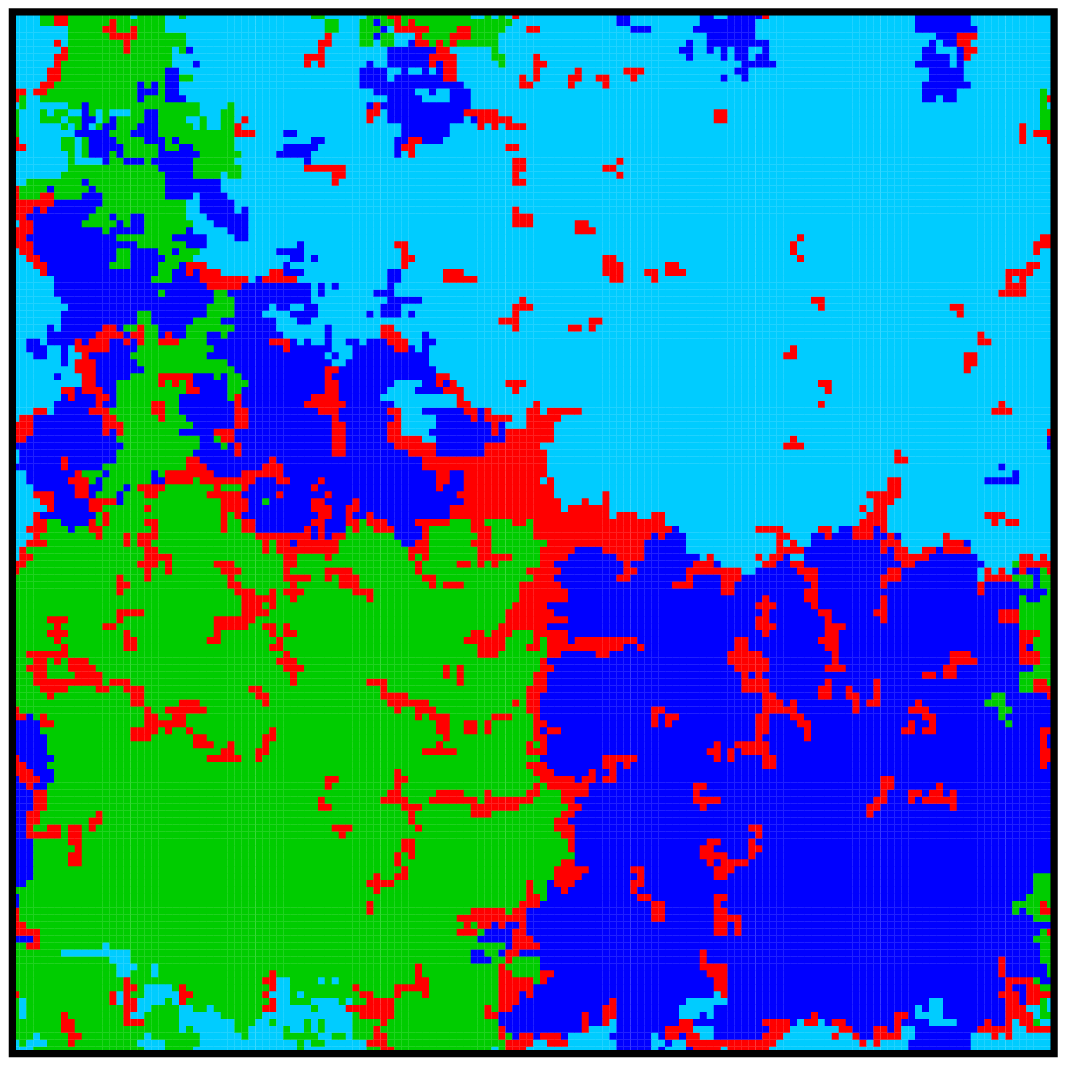,width=4.6cm}}
\centerline{\epsfig{file=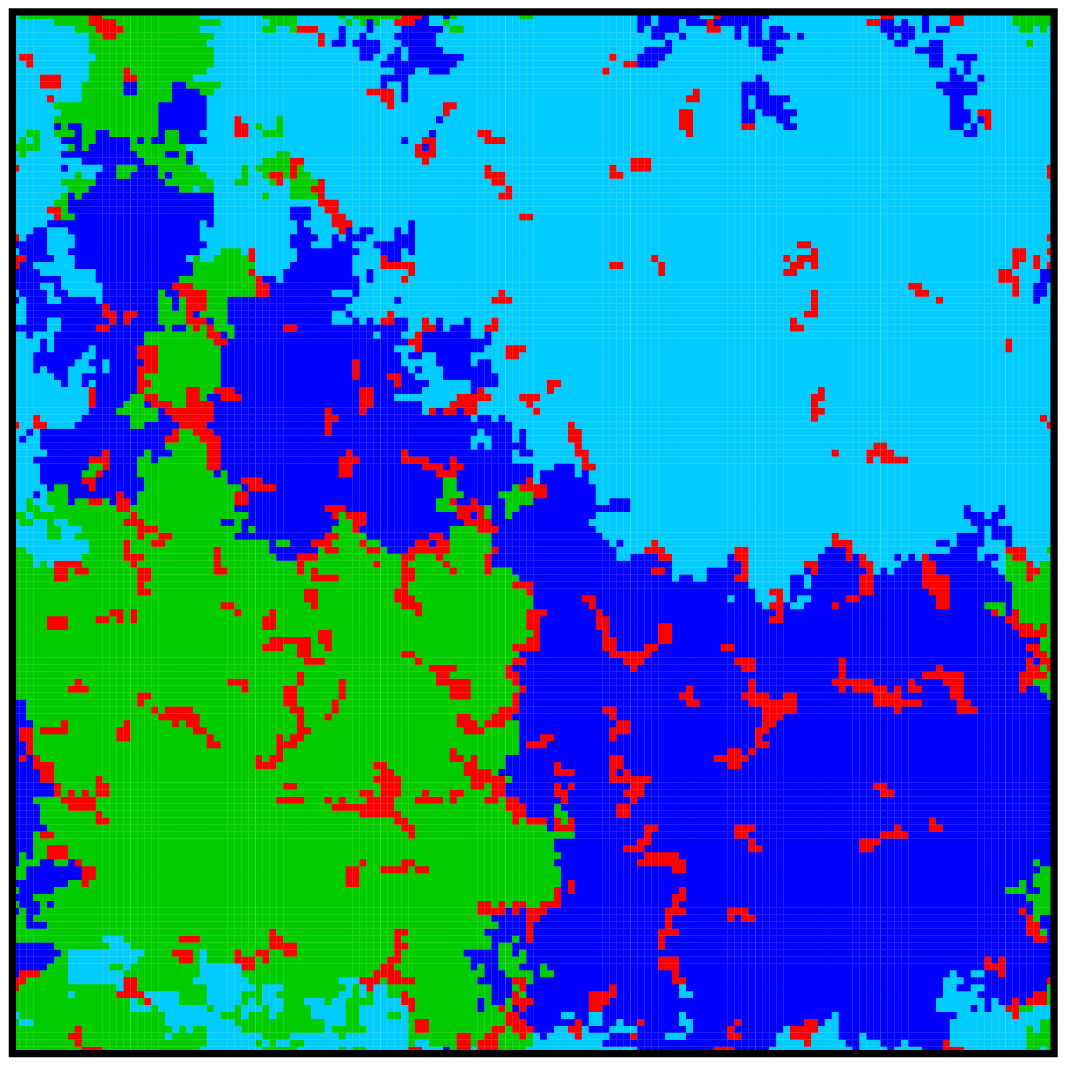,width=4.6cm}\epsfig{file=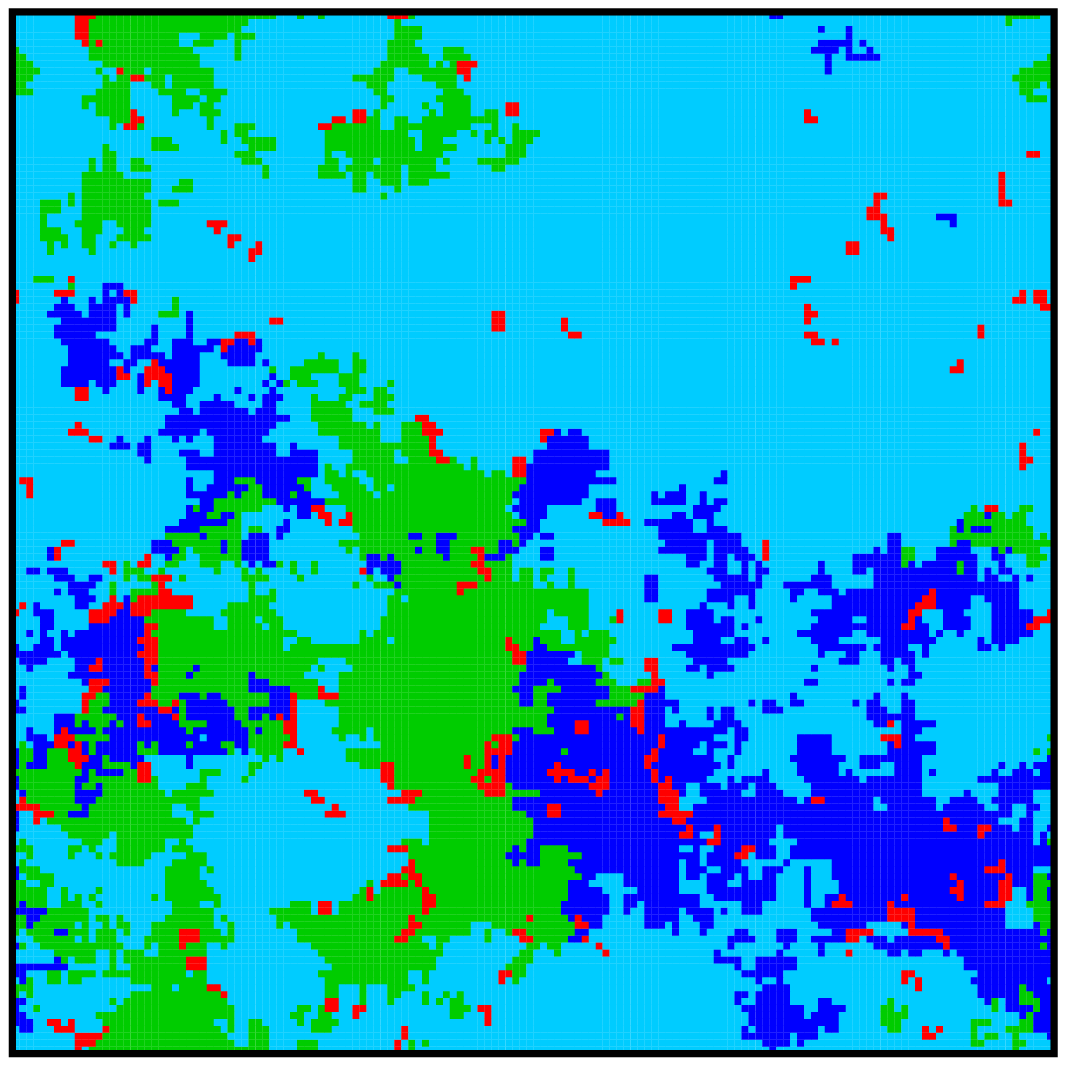,width=4.6cm}\epsfig{file=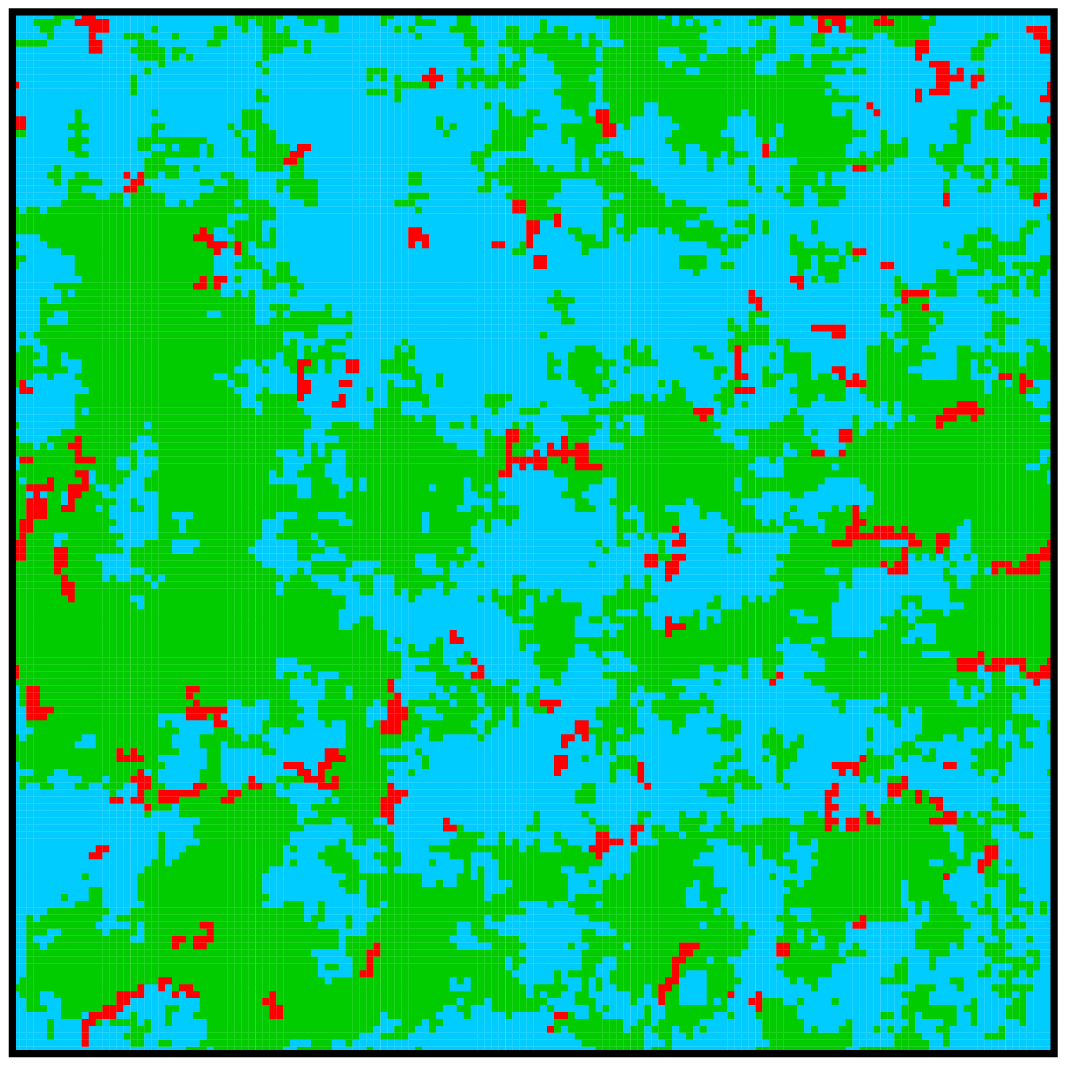,width=4.6cm}}
\caption{\label{r4_5} Snapshots of the square lattice, showing a characteristic evolution from a prepared initial state, as obtained for $\gamma=0.0015$, $\beta=0.1$ and $r=4.5$. Strategies $D$, $P$, $R$ and $B$ are depicted red, green, dark blue and light blue, respectively (the same colors were used in Fig.~\ref{r4_5_cross}). Time runs from the top left panel towards the bottom right panel at $0$, $100$, $210$, $300$, $1130$ and $8700$ Monte Carlo steps, respectively. At $0$ MCS the game is initiated from a prepared initial state (the upper left domain is a mixture of strategies $P$, $R$ and $B$) to allow the usage of a very small system size ($L=150$) that still allows to infer strategy configurations in sufficient detail. At $100$ MCS $D$ (red) percolate slightly into green, light blue and dark blue clusters, indicating that all three cooperative strategies could form a two-strategy phase with defectors. At $210$ MCS the borders of the three mentioned two-strategy phases ($D+P$, $D+R$, $D+B$) meet, and at $300$ MCS it can be observed that strategy $B$ (light blue) is capable to invade into the $D+P$ phase while strategy $R$ is unable to do the same. While $R$ and $B$ are neutral in the absence of $D$, in direct competition against $D$ the strategy $B$ is more effective and thus continues to crowd out strategy $R$, as can be inferred at $1130$ MCS. Interestingly, strategy $P$ has a small advantage over strategy $B$ because the former can spare the cost of rewarding. Yet this advantage is sufficient for $P$ to survive in the bulk of strategy $B$. The last panel, taken at $8700$ MCS, depicts a typical stationary pattern where strategies $D$, $P$ and $B$ coexist to form the $D+P+B$ phase.}
\end{figure*}

Systematic Monte Carlo simulations are performed to reveal phase diagrams for two representative values of the synergy factor $r$. In the absence of reward and punishment, cooperators survive only if $r>3.74$, and they are able to defeat defectors completely for $r>5.49$ \cite{szolnoki_pre09c}. Taking these as benchmark values, we focus on $r=4.5$ and $r=2.5$. For $r=4.5$ cooperators are able to coexist with defectors without support from additional strategies, solely on the basis of network reciprocity. This can thus be considered as lenient conditions for the evolution of public cooperation. For $r=2.5$, on the other hand, cooperators are unable to survive in the absence of reward or punishment, and this are hence adverse conditions for cooperative behavior to prevail. For both values of $r$, we determine the stationary fractions of strategies when varying the reward/fine $\beta$ and the cost $\gamma$. The transition points and the type of phase transitions are identified from Monte Carlo data collected with a sufficiently high accuracy (and frequency) in the close vicinity of the transition points. Finally, the phase boundaries, separating different stable solutions, are plotted in the full $\beta-\gamma$ phase diagrams. The obtained quantitative results are discussed in detail in the following two subsections.

\subsection{Synergy factor $r=4.5$}

The phase diagram depicted in Fig.~\ref{phd_r4_5} suggests that at such a high value of $r$ the far more effective action to outperform defectors is punishment rather than rewarding. The pure (or absorbing because the applied dynamical rule leaves the phase unchanged once the system arrives there) $D$ phase in the upper left corner of the $\beta-\gamma$ plane first gives way to the mixed $D+P$ phase, and subsequently to the pure $P$ phase as the fine (cost) increases (decreases). Only if the cost is negligible and the fine/reward is large are rewarding strategies able to survive. In this case defectors die out very soon, and from there on strategies $R$ and $B$ become equivalent since there is nobody left to punish. For the same reason strategy $P$ transforms to that of ordinary cooperation. Accordingly, strategies $R$ and $B$ are able to coexist alongside strategy $P$ as long as the cost of rewarding is sufficiently small to offset the second-order free-riding (because $P$ do not contribute to rewarding other cooperators). The phase is denoted appropriately as $P+(RB)$ in the lower right corner of Fig.~\ref{phd_r4_5}.

Yet Fig.~\ref{phd_r4_5} fails to convey the full story behind the depicted phase diagram. For very low values of $\gamma$ ($\sim 10^3$ smaller than $\beta$), the studies spatial public goods game reveals its true potential to yield rich dynamical behavior that gives rise to a truly intricate phase diagram. As can be observed in Fig.~\ref{phd_r4_5_enlarged}, no less than seven successive phase transitions can occur upon varying a single parameter (increasing $\beta$ at a fixed value of $\gamma$). In addition to the pure $P$ phase, we can observe two-strategy $D+P$, $D+B$ and $P+(RB)$ (note that here $R$ and $B$ are equivalent strategies) phases, three strategy $D+P+B$ and $D+P+R$ phases, and even the four-strategy $D+P+R+B$ phase. While the majority of phase transitions is continuous, the $D+P \to D+B$ phase transition is discontinuous due to an indirect territorial competition (see \cite{helbing_ploscb10, szolnoki_jtb13} for further examples of this phenomenon) between strategies $P$ and $B$. The two compete independently against the defectors, and the victor is determined by whoever is more effective. The nature of the other phase transitions is illustrated quantitatively in Fig.~\ref{r4_5_cross}, which shows a cross-section across $\beta$ for the most interesting value of $\gamma$.

\begin{figure}
\centerline{\epsfig{file=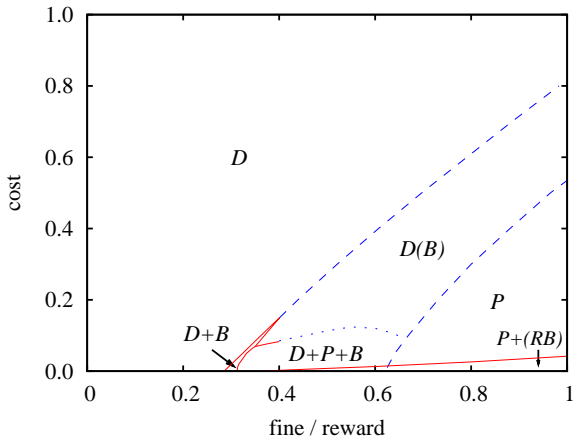,width=8.5cm}}
\caption{\label{phd_r2_5} Full $\beta-\gamma$ phase diagram, as obtained for $r=2.5$. Solid red lines denote continuous phase transitions, while dashed blue lines denote discontinuous phase transitions. More precisely, the dashed blue line between phases $D$ and $D(B)$ indicates that there would be a discontinuous phase transition if only $D$ and $B$ strategies were initially present in the system (in other words, the frontier of dominance between strategies $D$ and $B$ is delineated). The same holds for the dashed blue line separating phases $D(B)$ and $P$. The three-strategy $D+P+B$ phase is separated with a dotted blue line to emphasize that there are two very different ways in which this solution can give way to the $D(B)$ phase. In particular, at smaller fines the transition is continuous because the average fraction of strategy $P$ gradually decays to zero (see Fig.~\ref{r2_5_F0_370} for a quantitative insight). At larger fines, however, the averages of all three strategies remain finite, but the amplitude of oscillations diverges regardless of the system size (see Fig.~\ref{scaling} for details), which ultimately results in an abrupt termination of cyclic dominance between the three strategies (see Fig.~\ref{r2_5_F0_550}). The notation of the $P+(RB)$ phase at the lower right corner of the phase diagram has the same meaning as described in the caption of Fig.~\ref{phd_r4_5}.}
\end{figure}

Despite the complexity of solutions, the relevance of the presented results for the main question addressed in this study is quickly revealed. The dashed blue line in Fig.~\ref{phd_r4_5_enlarged}, marking the discontinuous $D+P \to D+B$ phase transition, conveys directly that the combined strategy $B$ is more effective than the elementary strategy $P$ only if $\beta$ increases [if conditions for rewarding and sanctioning become more lenient (the two actions become less costly)], and this only when the costs are already negligible ($\sim 10^3$ smaller than the administered rewards and fines). Accordingly, we conclude that, at least for high values of the synergy factor $r$, there are no notable evolutionary advantages associated by correlating positive (rewarding) and negative (punishment) reciprocity in a single strategy. This agrees with the empirical data presented by Yamagishi et al. \cite{yamagishi_pnas12} and Egloff et al. \cite{egloff_pnas13},
who failed to observe the same correlation in human experiments. On the other hand, it should not be overlooked that the combined strategy $B$ \textit{is} viable and that it does convey some advantages (albeit in very narrow and rather unrealistic parameter regions), which thus also lends some support to the strong reciprocity model \cite{fehr_hn02}.

\begin{figure}
\centerline{\epsfig{file=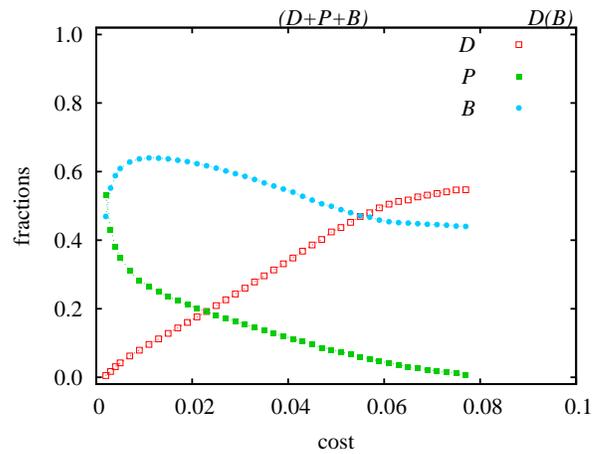,width=8.5cm}}
\caption{\label{r2_5_F0_370} Cross-section of the phase diagram depicted in Fig.~\ref{phd_r2_5}, as obtained for $\beta=0.37$. Depicted are stationary fractions of the four competing strategies in dependence on $\gamma$. Stable solutions are denoted along the top axis. In this cross-section the $D+P+B \to D(B)$ phase transition is continuous because the average fraction of strategy $P$ gradually decays to zero as $\gamma$ increases.}
\end{figure}

To demonstrate just how the combined strategy $B$ may survive, we show in Fig.~\ref{r4_5} a series of snapshots from a prepared initial state (applied solely to allow the usage of a relatively small system size), which eventually evolves towards the three-strategy $D+P+B$ phase. Based on this example, it could be argued that adopting strategy $B$ does in fact confer an advantage over strategy $R$, which succumbs to the evolutionary pressure stemming from the three surviving strategies. However, as can be observed at a glance from the depicted phase diagrams presented in Figs.~\ref{phd_r4_5} and \ref{phd_r4_5_enlarged}, this is limited to a very narrow and specific parameter range, which is practically invisible at normal resolution (see Fig.~\ref{phd_r4_5}). In addition, we emphasize that strategy $B$ is slightly less effective than strategy $P$ (see Fig.~\ref{r4_5_cross} for the stationary fractions of the two strategies). Thus, although the combined strategy $B$ might appear as a good choice in some of the parameter regions within Fig.~\ref{phd_r4_5_enlarged}, it is still second best to the elementary strategy $P$ adopting solely punishment.

\subsection{Synergy factor $r=2.5$}

\begin{figure}
\centerline{\epsfig{file=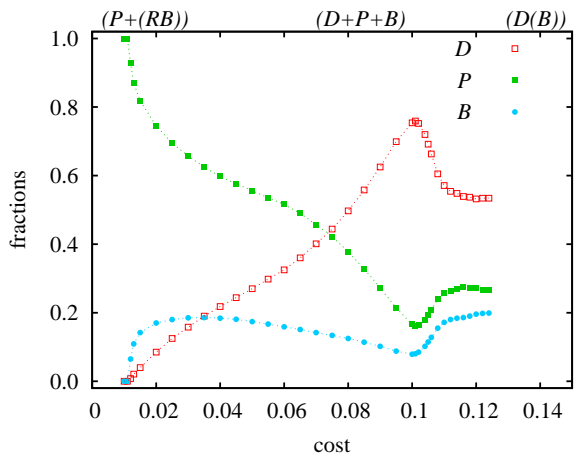,width=8.5cm}}
\caption{\label{r2_5_F0_550} Cross-section of the phase diagram depicted in Fig.~\ref{phd_r2_5}, as obtained for $\beta=0.55$. Depicted are stationary fractions of the four competing strategies in dependence on $\gamma$. Stable solutions are denoted along the top axis. Unlike in Fig.~\ref{r2_5_F0_370}, here the $D+P+B \to D(B)$ phase transition is discontinuous because the amplitude of oscillations diverges independently of the system size (see Fig.~\ref{scaling} for details) as $\gamma$ increases.}
\end{figure}

If the conditions for the evolution of public cooperation become harsh, as is the case for $r=2.5$, the relations between the competing strategies change quite significantly. The phase diagram presented in Fig.~\ref{phd_r2_5} reveals that, besides the expected extension of the pure $D$ phase, the parameter region where strategy $B$ can survive also becomes larger. Furthermore, there is a significant change in the nature of phase transitions. Unlike at $r=4.5$ (see Fig.~\ref{phd_r4_5}), here discontinuous phase transitions dominate, which has to do with the spontaneous emergence of cyclic dominance \cite{reichenbach_n07, reichenbach_prl07, mobilia_jtb10, mobilia_epl11} between strategies $D$, $P$ and $B$. In particular, within the three-strategy $D+P+B$ phase strategy $D$ outperforms strategy $P$, strategy $P$ outperforms strategy $B$, while strategy $B$ again outperforms strategy $D$. It is important to note that at $r=4.5$ the stability of none of the three-strategy phases, and also not of the four-strategy phase, has been due to cyclic dominance. Instead, as Fig.~\ref{r4_5} illustrates, there the stability was warranted by the stable coexistence of the strategies, rather than by oscillations that are brought about by cyclic dominance.

As was frequently the case before \cite{wu_zx_pre05, szolnoki_pre11, szolnoki_pre11b, szolnoki_prl12}, here too the spontaneous emergence of cyclic dominance brings with it fascinating dynamical processes that are driven by pattern formation, by means of which the phase may terminate. Figures~\ref{r2_5_F0_370} and \ref{r2_5_F0_550} feature two characteristic cross-sections of the phase diagram presented in Fig.~\ref{phd_r2_5}, which reveal two qualitatively very different ways for the $D+P+B$ cyclic dominance phase to give way to the $D(B)$ phase [here $D(B)$ indicates that either a pure $D$ or a pure $B$ phase can be the final state if starting from random initial conditions]. The process depicted in Fig.~\ref{r2_5_F0_370} is relatively straightforward. Here the average fractions of strategies $P$ and $B$ decay due to the increasing cost $\gamma$, which ultimately results in the vanishing average value of the fraction of strategy $P$. The closed cycle of dominance is therefore interrupted and the $D+P+B$ phase terminates.

\begin{figure}
\centerline{\epsfig{file=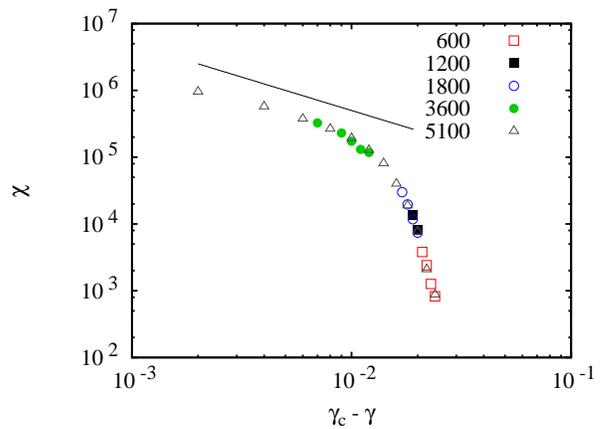,width=8.5cm}}
\caption{\label{scaling} Fluctuations of the amplitude of oscillations $\chi$ in dependence on the vicinity to the critical value of the cost $\gamma_c=0.1242(6)$ for $\beta=0.55$ and different system sizes, as indicated in the legend. The slope of the power-law exponent (solid line) is $1$, indicating divergent fluctuation as $\gamma$ approaches the critical value.}
\end{figure}

\begin{figure}[b]
\centerline{\epsfig{file=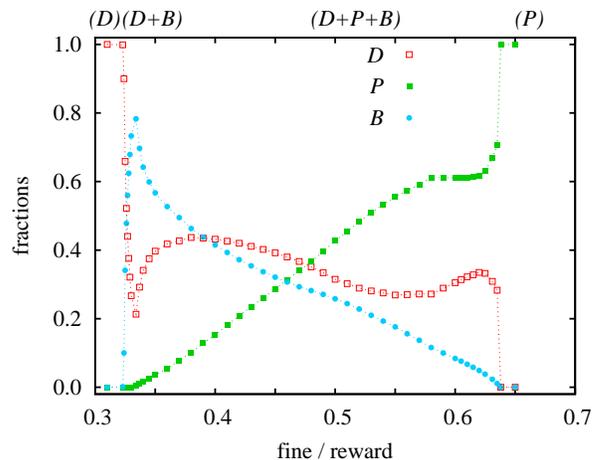,width=8.5cm}}
\caption{\label{r2_5_cost0_05} Cross-section of the phase diagram depicted in Fig.~\ref{phd_r2_5}, as obtained for $\gamma=0.05$. Depicted are stationary fractions of the four competing strategies in dependence on $\beta$. Stable solutions are denoted along the top axis. In this cross-section the pure $D$ phase first gives was to a very narrow interval where the two-strategy $D+B$ phase precedes the spontaneous emergence of cyclic dominance between the strategies $D$, $P$ and $B$. The last (from left to right) $D+P+B \to P$ phase transition is qualitatively similar to the $D+P+B \to D(B)$ phase transition depicted in Fig.~\ref{r2_5_F0_370}, only that here it is the fraction of strategy $B$ that decays to zero as $\beta$ increases, and which thus interrupts the closed cycle of dominance (in Fig.~\ref{r2_5_F0_370} it is the fraction of strategy $P$ as $\gamma$ increases that has the same effect).}
\end{figure}

\begin{figure*}
\centerline{\epsfig{file=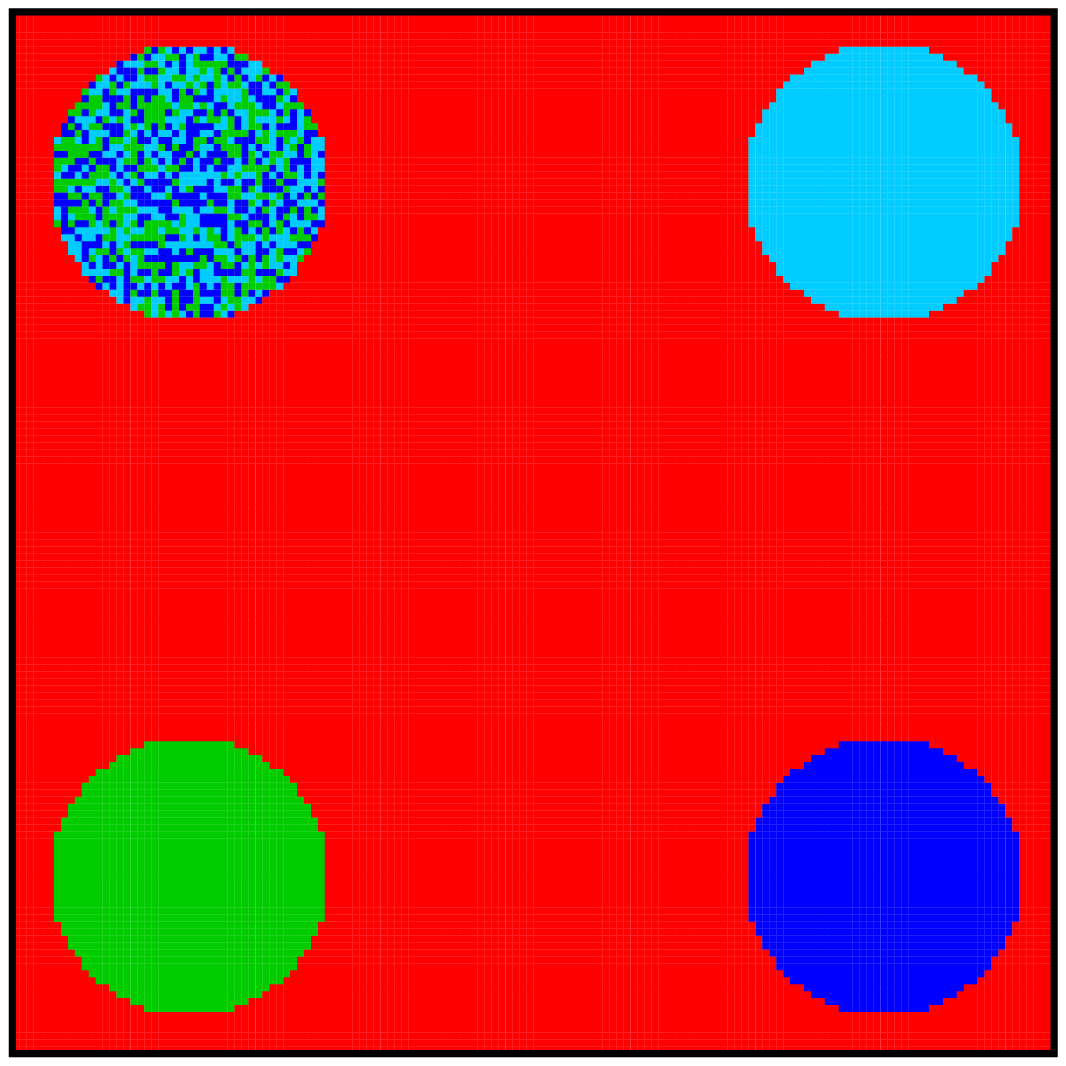,width=4.6cm}\epsfig{file=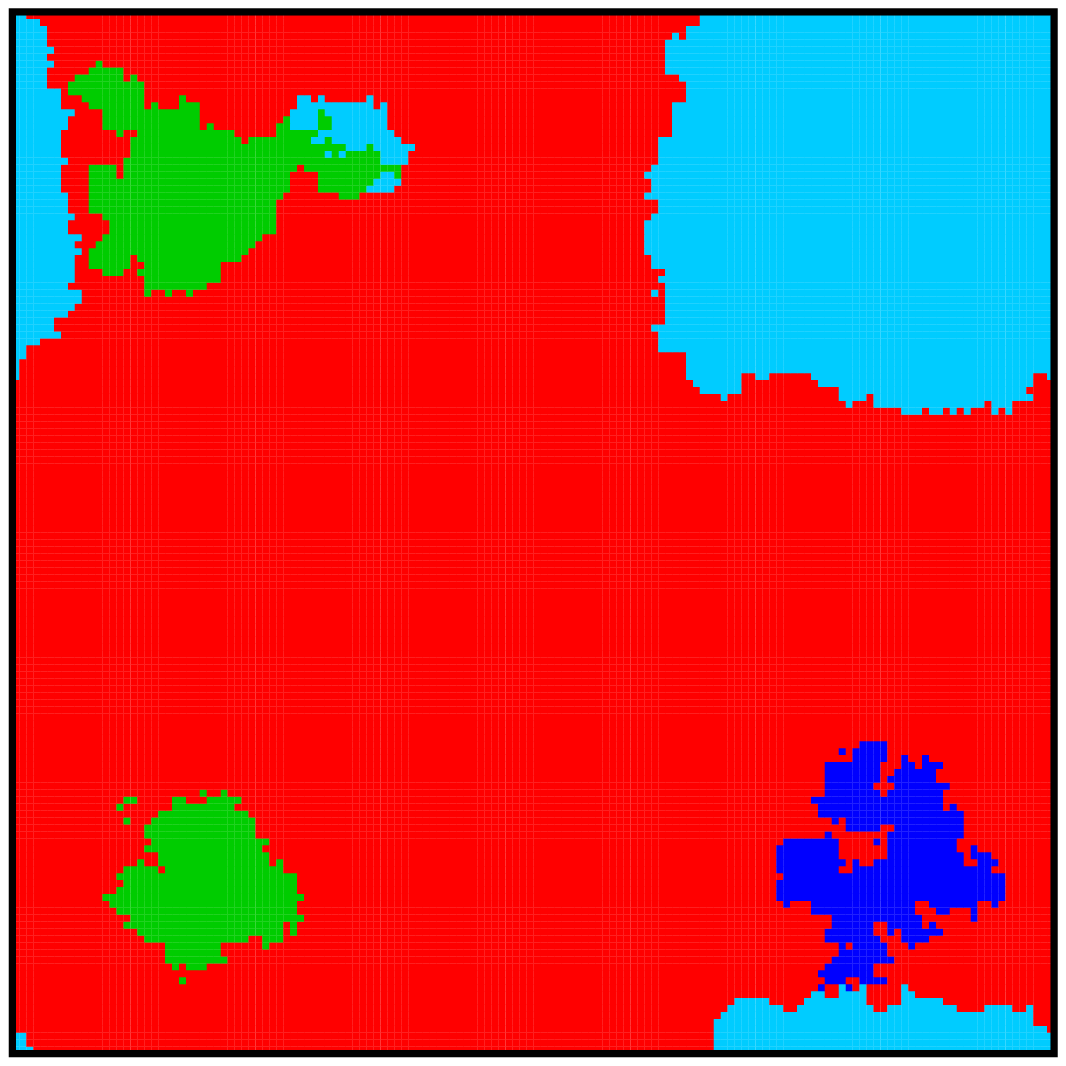,width=4.6cm}\epsfig{file=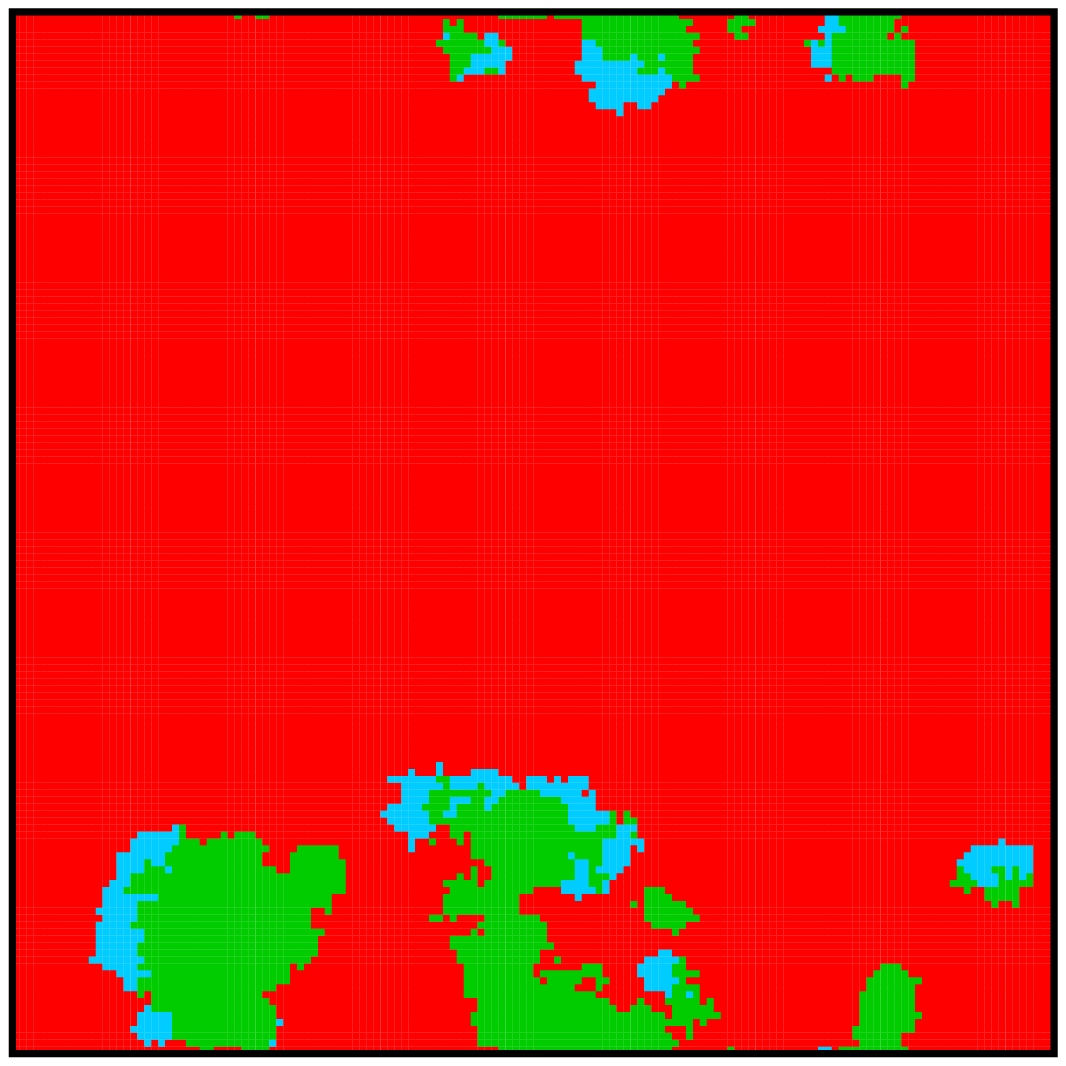,width=4.6cm}}
\centerline{\epsfig{file=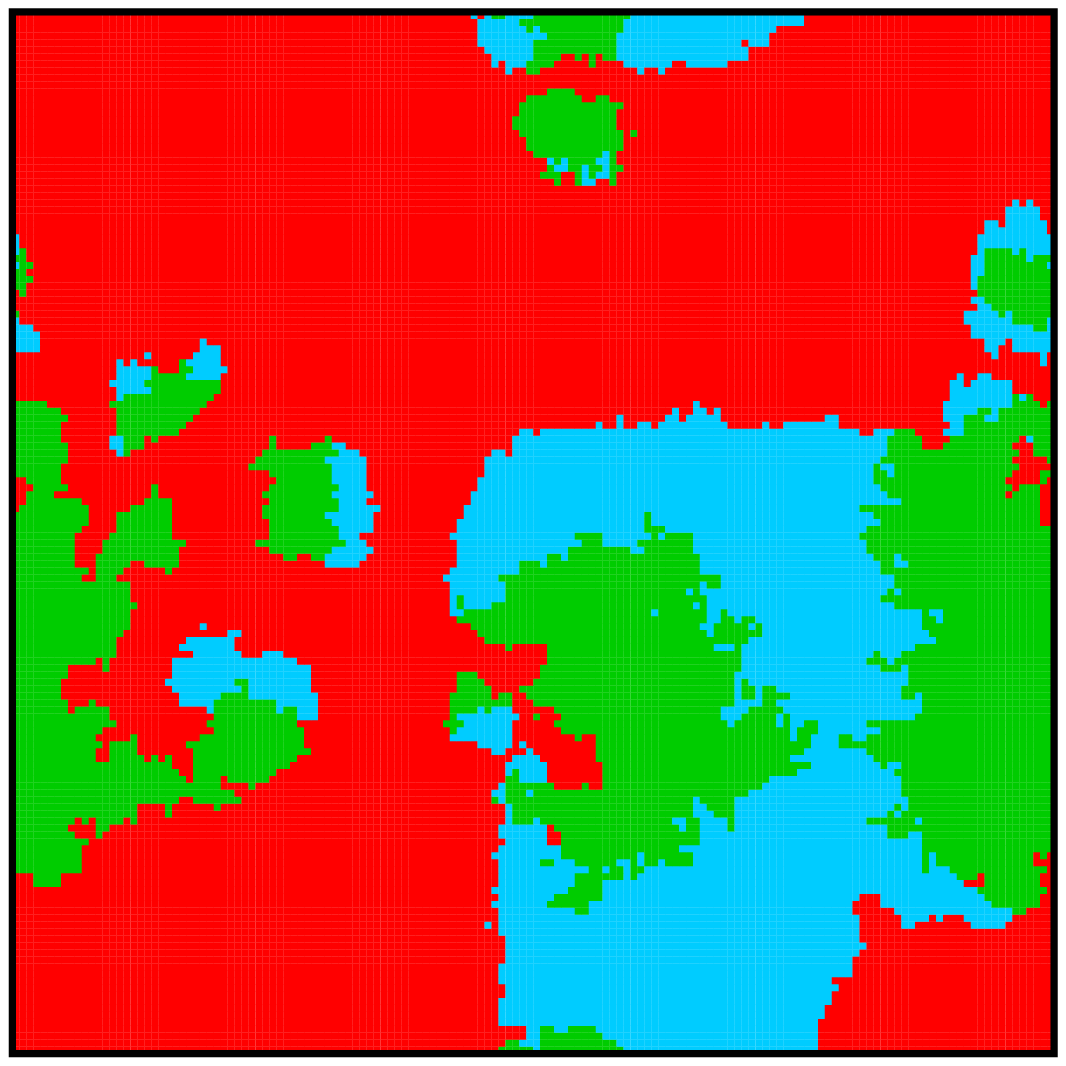,width=4.6cm}\epsfig{file=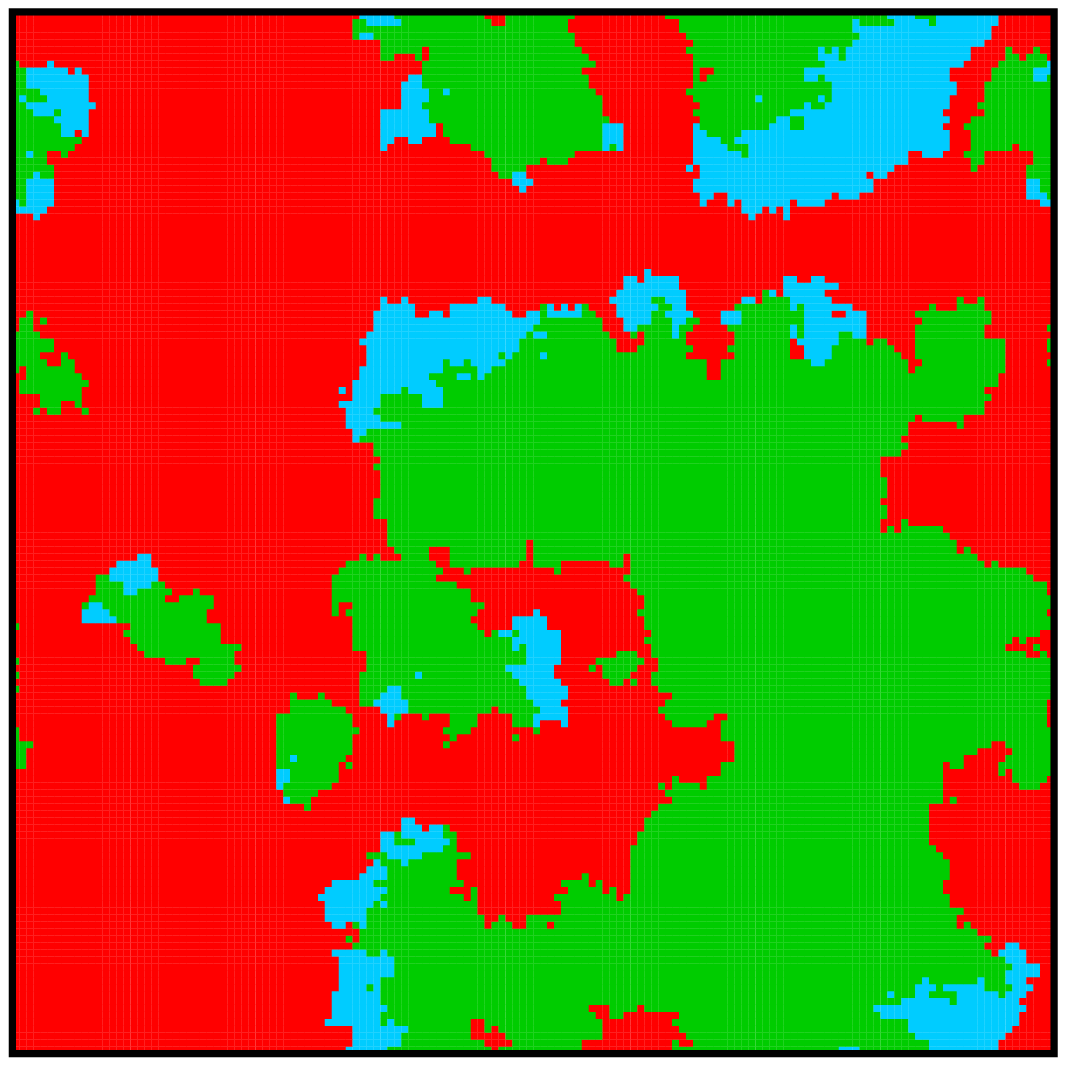,width=4.6cm}\epsfig{file=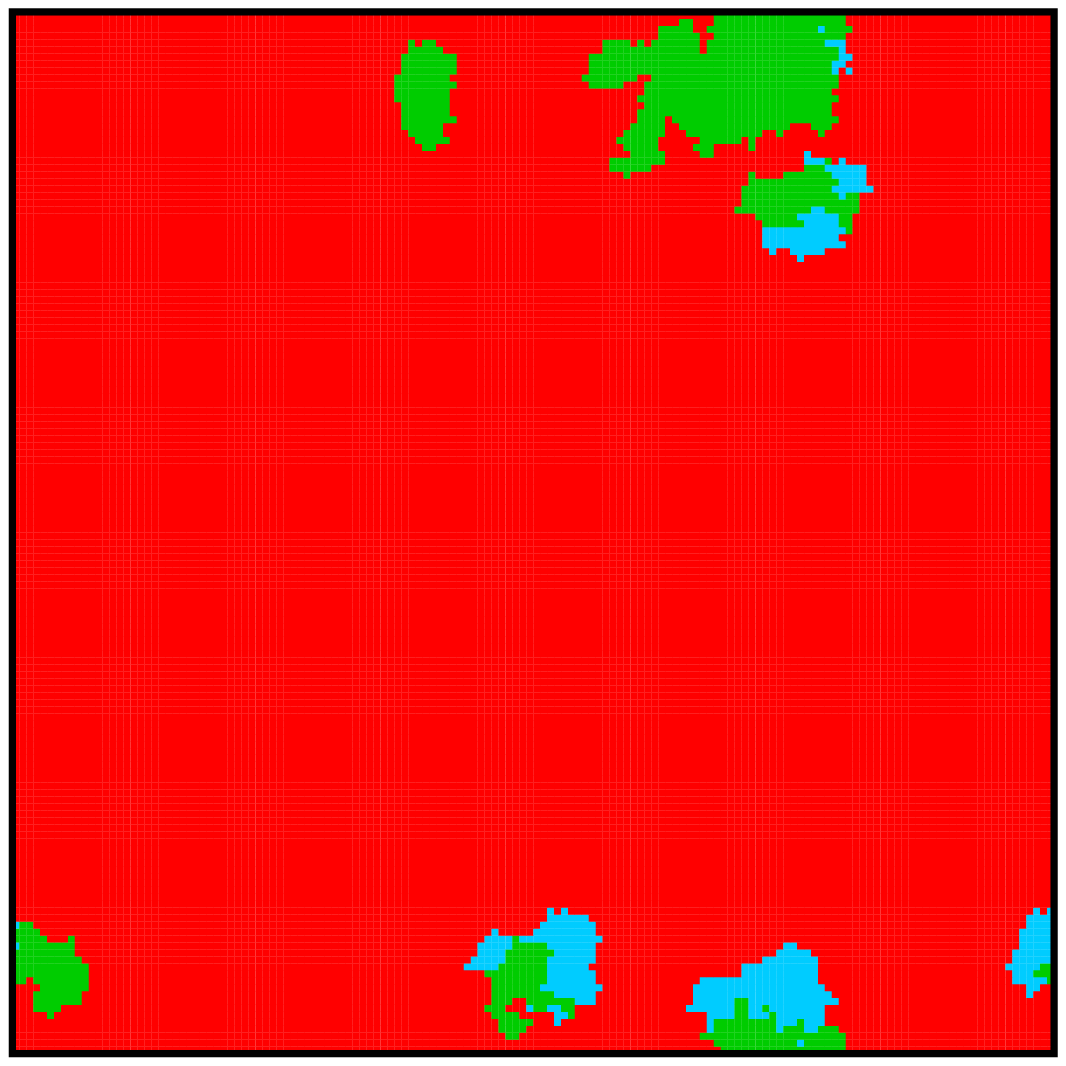,width=4.6cm}}
\caption{\label{r2_5} Snapshots of the square lattice, showing a characteristic evolution from a prepared initial state, as obtained for $\gamma=0.1$, $\beta=0.55$, and $r=2.5$. Strategies $D$, $P$, $R$ and $B$ are encoded with the same color as used in Fig.~\ref{r4_5}. Time runs from the top left panel towards the bottom right panel at $0$, $100$, $1800$, $4100$, $4240$ and $4610$ Monte Carlo steps, respectively. At $0$ MCS the game is initiated from the same prepared initial state, and for the same reason, as described in the caption of Fig.~\ref{r4_5}. At $100$ MCS, it can be observed that strategies $P$ (green) and $R$ (dark blue) are both weaker than strategy $D$ (red), and accordingly their isolated islands shrink. Conversely, the combined strategy $B$ is more effective when competing alone against the defectors, and thus the light blue island grows. Moreover, in the absence of defectors the strategy $P$ can exploit rewarding strategies and spread fast in the bulk of the mixed domain (left upper circle). It is also worth pointing out that the $B$ domain would grow endlessly in the sea of defectors if it would not meet the elementary strategy $P$ that is able to exploit it. At $1800$ MCS the final solution is practically formed, and from here on traveling waves dominate the spatial grid. At $4100$ MCS it can be observed that the strategy $B$ can spread towards strategy $D$, and based on this it may control a significant portion of the lattice for a short period of time. At $4240$ MCS, however, the strategy $P$ can easily invade the bulk of the $B$ domain, but in the absence of the latter it itself becomes vulnerable against the defectors. This cycle of dominance is repeated from $4610$ MCS onwards, which is a very similar configuration as the one at $1800$ MCS. Naturally, the oscillations become more intense as we approach the edge of $D+P+B$ phase in Fig.~\ref{phd_r2_5}, and the evolution can easily terminate in an absorbing phase if the system size is not sufficiently large.}
\end{figure*}

The situation for $\beta=0.55$ is much more peculiar and interesting. As results presented in Fig.~\ref{r2_5_F0_550} demonstrate, here the average values of all three strategies remain finite. Hence, the termination of the $D+P+B$ phase must have a different origin than at $\beta=0.37$ presented in Fig.~\ref{r2_5_F0_370}. In fact, for $\beta=0.55$ it is the amplitude of oscillations that increases with increasing values of $\gamma$. And it is the increase in the amplitude that ultimately results in a uniform absorbing phase regardless of the system size. At this point it is crucial to emphasize that the increase of the amplitude of oscillation is not a finite-size effect. Although in spatial systems with cyclic dominance it is typical to observe oscillations with increasingly smaller amplitude as the system size is increased, this does not hold in the present case. To demonstrate this, we measure the fluctuations in the stationary state according to
\begin{equation}
\chi= {L^2 \over M} \sum_{t_i = 1} ^M \left \langle \left( \rho_D(t_i)- \overline{\rho_D} \right)^2 \right \rangle \,
\label{eq:chi}
\end{equation}
where $\overline{\rho_D}$ is the average value of the fraction of defectors (a similar quantity can be calculated for the other strategies as well). As Fig.~\ref{scaling} shows, the scaled quantity $\chi$ is size-independent, thus indicating a divergent fluctuation as $\gamma$ approaches the critical value. The three-strategy $D+P+B$ phase is therefore unable to exist beyond this value despite the fact that the average fractions of all three strategies are far from zero. Instead, the phase terminates via a discontinuous phase transition towards the $D(B)$ phase, as depicted in Fig.~\ref{phd_r2_5}. Notably, within the $D(B)$ phase either the pure $D$ or the pure $B$ phase can be the final state, depending on which strategy dies out first.

With this, however, we have not yet covered all the details of the phase diagram presented in Fig.~\ref{phd_r2_5}. In addition, there are namely the same pure $P$ and two-strategy $P+(RB)$ phases observable that we have already reported above for $r=4.5$ (see Fig.~\ref{phd_r4_5}), only that at $r=2.5$ they are shifted further towards higher values of $\beta$. This is understandable, given that the lesser support for public cooperation due to a lower value of the synergy factor needs to be offset by higher fines and rewards. Moreover, we must not overlook the existence, albeit a very subtle one, of the two-strategy $D+B$ phase, the emergence of which is quantitatively described in the cross-section presented in Fig.~\ref{r2_5_cost0_05}. This is the only stable solution where solely the combined strategy $B$ coexists with defectors, and where thus the correlation of negative and positive reciprocity truly outperforms elementary strategies $P$ and $R$. As in all the previously outlined cases, however, in this case too this advantage is minute and limited to a very narrow region in the phase diagram.

In general, the harsher conditions for the evolution of public cooperation lend more support for the combined strategy to survive, as indeed the regions on the $\beta-\gamma$ parameter plane where $B$ can prevail become quite extensive at smaller values of the synergy factor $r$. This extends the credibility of the strong reciprocity model, and it indicates that, if at all, the evolutionary advantages of correlated positive and negative reciprocity ought to manifest clearer under extreme adversity. In future experiments, it may thus be worthwhile working towards such conditions if the goal is to possibly discern some more actual advantages of correlated reciprocities, and to thus further support the assumptions of the strong reciprocity theory also with empirical data. A warning to end the presentation of results is, however, in order. As the series of final snapshots presented in Fig.~\ref{r2_5} clearly demonstrates (and to no lesser extent also the series of snapshots presented in Fig.~\ref{r4_5}), conditions for pattern formation and complex strategic configurations need to be given for the subtle solutions, here identified by means of extensive and systematic Monte Carlo simulations, to emerge and be stable. Such conditions appear to be very difficult to achieve in experiments with humans, which is why efforts towards large-scale implementations, as recently reported in \cite{gracia-lazaro_srep12, gracia-lazaro_pnas12}, are very encouraging and certainly worth developing further in the future.

\section{Discussion}

Our goal in the present paper was to determine whether there are evolutionary advantages associated by correlating positive and negative reciprocity in a single strategy, as opposed to adopting solely reward or punishment as an elementary strategy. Systematic Monte Carlo simulations have revealed that, regardless of the synergy factor governing the public goods game, elementary strategies, and punishment in particular, are in general significantly more effective in deterring defection than the combined strategy. Although there exist narrow and rather unrealistic parameter regions where the correlation of positive and negative reciprocity can outperform a particular elementary strategy, these advantages are highly unlikely to play a role in human experiments, and they also frequently come second to the evolutionary success that is warranted by punishment alone under the same conditions. The presented results thus lend support to the empirical data published in \cite{yamagishi_pnas12, egloff_pnas13}, which fail to support the central assumption of the strong reciprocity model that negative and positive reciprocity are correlated.

The studied four-strategy spatial public goods game gives rise to fascinating evolutionary outcomes that are separated by continuous and discontinuous phase transitions. We have demonstrated, for example, that indirect territorial competition may lend some credibility to the combined strategy, as the latter is sometimes more effective against the defectors than solely rewarding. In special parameter regions, the combination of positive and negative reciprocity can thus crowd out cooperators that reward other cooperators. Under the same conditions, however, cooperators that punish but do not reward can be more effective still, so overall it is difficult to argue in favor of choosing the combined strategy over an elementary one. Moreover, stationary solutions that are governed by indirect territorial competition terminate suddenly via discontinuous phase transitions, and accordingly, they are difficult to identify and are unlikely to seriously challenge conclusions based on empirical data.

For low synergy factors, we have shown that the spontaneous emergence of cyclic dominance between strategies $D$, $P$ and $B$ is also a possible solution of the system, and indeed it significantly extends the parameter region where the correlation of positive and negative reciprocity is viable. Within the cyclic phase defectors outperform punishers, punishers outperform the combined strategy, and the combined strategy is able to invade defectors, thereby closing the loop of dominance. In this case, it can again be argued in favor of the combined strategy over solely rewarding, but since the remaining three strategies become spontaneously entailed in a cycle of dominance, the advantage warranted by the correlation of negative and positive reciprocity is indirect and circumstantial at best. Furthermore, we have demonstrated that the cyclic dominance can terminate in very different ways. Either the average fraction of one strategy vanishes, or, more intriguingly, the amplitude of oscillations diverges in a system-size independent manner. Thus, although the average fractions of all three strategies are far from zero, the cyclic dominance phase may end abruptly via a discontinuous phase transition. Although phenomena like indirect territorial competition, cyclic dominance, divergent fluctuations of the amplitude of oscillations, as well as previously reported critical phenomena in evolutionary games \cite{hernandez_epjb11}, self-organized adaptation \cite{timme_np10, lee_s_prl11} and in-group favoritism \cite{fu_srep12}, are all of significant interest to physicist, we emphasize that they would likely require massive efforts to be observed in human experiments. Nevertheless, recent large-scale attempts in this direction promise exciting times ahead \cite{gracia-lazaro_srep12, gracia-lazaro_pnas12}.

Lastly, it remains to emphasize that punishment is the elementary strategy that is definitively more effective than the combined strategy, while rewarding is not necessarily so. However, rewarding can be made much more potent if rewards are administered not to all cooperators, but only to those who themselves reward others. In this case, rewarding can completely outperform punishment at low $\gamma$ and high $\beta$ values, while the situation reverses only if the costs become relatively high compared to the rewards and fines. Yet in this modified scenario, the act of punishing yields no extra advantages, and in general the strategy $B$ can survive only when the strategy $R$ can survive too. Therefore, even under such altered, rewarding-friendly conditions, there are still no notable evolutionary advantages to be gained by adopting a strategy that combines both positive and negative reciprocity. With this conclusion, we hope that our study will inspire further research aimed at investigating the role of correlated strategies in evolutionary games, and we also hope that more experimental work will be carried out to clarify their role by the evolution of human cooperation.

\begin{acknowledgments}
This research was supported by the Hungarian National Research Fund (Grant K-101490) and the Slovenian Research Agency (Grant J1-4055).
\end{acknowledgments}

\section*{}
\section*{Popular summary}

Widespread cooperation among unrelated individuals distinguishes humans markedly from other species. The origins of our remarkable other-regarding abilities have been associated with between-group conflicts as well as with alloparental care and provisioning for someone else's young, all of which were pressing challenges that during the Paleolithic age and beyond could not be met by individual efforts. But in the absence of such challenges, what keeps us cooperating? Reciprocity is long considered an important piece of the puzzle. If someone is kind to us, we tend to be kind in return. Similarly, if someone is unfair or exploitative, we retaliate without much sympathy. And according to the strong reciprocity hypothesis, this positive and negative reciprocity are correlated to give us optimal evolutionary predispositions for the successful evolution of cooperation. But is this really true? Recent human experiments reject the hypothesis, and everyday experiences leave us with the impression that people either tend to reward success or punish wrongdoing, but seldom will they do both. We show how methods of statistical physics resolve the disagreement between theory and observations.

We perform Monte Carlo simulations of the public goods game, where besides defectors, cooperators have the option to reward, punish, or do both. The question we focus on is whether doing both confers an evolutionary advantage to individuals adopting such a correlated strategy. The evolutionary game gives rise to indirect territorial competition, spontaneous emergence of cyclic dominance, as well as divergent fluctuations of oscillations that terminate in an absorbing phase. All these are fascinatingly complex solutions that highlight the importance of collective behavior, pattern formation, and structure in human cooperation. Yet despite the richness of evolutionary outcomes, the correlated strategy can survive only in very narrow and rather unrealistic parameter regions. Rewarding or punishing only is far superior and more effective in deterring defection. Presented results thus lend support to the outcome of recent human experiments, which fail to support the strong reciprocity hypothesis that positive and negative reciprocity are correlated.

Our study demonstrates the usage of statistical physics in evolutionary games with correlated strategies, and we hope it will inspire more theoretical and experimental work aimed at clarifying their role by the evolution of human cooperation. The richness of the results and associated phase diagrams might also attract the attention of the broader physics community to be inspired by the complexity and beauty of the physics of non-equilibrium systems that address fundamental social and biological problems.

\end{document}